\documentclass[10pt]{iopart}

\usepackage{times}
\usepackage{colordvi}
\usepackage{graphicx}

\usepackage{iopams}
\usepackage{amssymb}

\def\de{\partial} 
\def\r{\hat{r}}

\begin{document}

\title[Binary black hole merger in the extreme mass ratio limit]{Binary black hole 
merger in the extreme mass ratio limit}

\author{Alessandro Nagar$^1$, Thibault Damour $^2$,  and Angelo Tartaglia$^1$ }

\address{$^1$ Dipartimento di Fisica, Politecnico di Torino, Corso Duca degli Abruzzi 24,
              10129 Torino, Italy and INFN, sez. di Torino, Via P.~Giuria 1,
              Torino, Italy}

\address{$^2$Institut des Hautes Etudes Scientifiques, 35 route de Chartres, 91440 Bures-sur-Yvette, France}

\begin{abstract}
  We discuss the   transition   from  quasi-circular  inspiral  to plunge of a 
  system of two nonrotating black holes of masses $m_1$  and  $m_2$ in the 
  extreme mass ratio limit $m_1m_2\ll (m_1+m_2)^2$. In the spirit of the Effective 
  One Body (EOB) approach to the general relativistic dynamics of binary systems, 
  the dynamics of the two black hole system is represented in terms of an effective 
  particle of mass $\mu\equiv m_1m_2/(m_1+m_2)$ moving in a (quasi-)Schwarzschild 
  background of mass $M\equiv m_1+m_2$ and submitted to an ${\cal O}(\mu)$ radiation 
  reaction force defined by Pad\'e resumming high-order Post-Newtonian results. We 
  then complete this approach by numerically computing, \`a la Regge-Wheeler-Zerilli, 
  the gravitational radiation emitted by such a particle. Several tests of the numerical 
  procedure are presented. We focus on gravitational waveforms and the related energy 
  and angular momentum losses. We view this work as a contribution to the matching 
  between analytical and numerical methods within an EOB-type framework.
\end{abstract}


\submitto{\CQG}



\section{Introduction}
\label{sec:intro}

The last months have witnessed a decisive advance in Numerical Relativity,
with different groups being able to simulate, for the first time and by using 
different techniques, the merger of two black holes of comparable masses 
(without or with initial spin)~\cite{pretorius05a,campanelli06a,baker06a,campanelli06b}. 
Since such binary black holes systems (of a total mass$\sim30 M_{\odot}$) 
are believed to be among the most promising sources of gravitational waves for 
the ground based detectors like LIGO and VIRGO, this breakthrough 
raises the hope to have, for the first time, a reliable estimate of the 
complete waveform by joining together Post-Newtonian (PN) 
and Numerical Relativity results. We recall that PN techniques have provided 
us with high-order results for describing the motion~\cite{DJS01,BDEF04} 
and radiation~\cite{BDEI04,blanchet_lrr} of binary systems, and that further techniques 
have been proposed for {\it resumming} the PN results~\cite{DIS,BD99,BD00}, 
thereby allowing an analytical description of the gravitational waveform 
emitted during the transition between inspiral and plunge, and even during 
the subsequent merger and ringdown phases. We now face the important task 
of constructing accurate {\it complete waveforms} by {\it matching} together 
the information contained in {\it Post-Newtonian} and {\it Numerical Relativity} 
results. We view the present work as a contribution towards this goal (for a 
recent first cut at this problem see~\cite{BCP06a,Baker_2006ha,Baker_2006kr}).

The work we present here belongs to a scientific lineage which started 
with the pioneering works of  Regge and Wheeler~\cite{RW57}, 
Zerilli~\cite{zerilli70}, Davis, Ruffini, Press and Price~\cite{DRPP}  
and Davis, Ruffini and Tiomno~\cite{DRT}. References~\cite{DRPP,DRT}
studied  the gravitational wave emission due to the {\it radial plunge} 
(from infinity) of a particle onto a Schwarzschild black hole. 
This was thought of as  a model for the head-on collision of two 
black holes in the extreme mass ratio limit. The gravitational wave 
energy spectrum~\cite{DRPP} and waveform~\cite{DRT} were obtained. 
Davis, Ruffini and Tiomno~\cite{DRT} pointed out that the first
part of the waveform could be described in terms essentially of 
a flat space ``quadrupole formula'' for a test particle following a
Schwarzschild dynamics (Ruffini-Wheeler approximation~\cite{RW71}),
while the last part of the waveform was dominated by exponentially
damped harmonic oscillations. These damped oscillations were 
interpreted by Press~\cite{press71} (see also Vishveshwara~\cite{vishu70})
as vibrational modes (now called quasi-normal modes)
of a Schwarzschild black hole. The perturbative formalism employed 
in these works has been later shown to be expressible in a gauge 
invariant manner~\cite{monc74,ST01,MP05,NR05}. The case of a particle 
plunging from infinity with {\it nonzero angular momentum} has been discussed 
by Detweiler and Szedenits~\cite{DS79} and Oohara and Nakamura~\cite{ON83} 
by means of the curvature perturbation formalism of Teukolsky~\cite{teuk73}. 
The radial plunge problem has been recently extended to a plunge from a 
finite distance~\cite{LP97,MP02}, with particular emphasis on the effect 
of the choice of initial data. 

However, none of the above works has studied the {\it transition} 
from the quasi-circular adiabatic inspiral phase to the plunge phase 
in extreme-mass-ratio binary black hole systems. The reason is that
the original Regge-Wheeler-Zerilli test-particle approach seems
to require the test particle to follow an exact {\it geodesic}
of the Schwarzschild background. Here we shall bypass this 
stumbling block by appealing to some results of PN theory.
In particular, the Effective One Body (EOB) approach to the general 
relativistic two body dynamics, which has been recently proposed to 
study the transition from inspiral to plunge in the comparable-mass 
case~\cite{BD99,BD00,DJS00,D01,BDC05,DG06}, describes the dynamics of 
a binary system in terms of two separate ingredients: (i) a Hamiltonian
$H_{\rm EOB}(M,\mu)$ describing the {\it conservative} part of the 
relative dynamics, and (ii) a non-Hamiltonian supplementary force 
${\cal F}_{\rm EOB}(M,\mu)$ approximately describing the reaction
to the loss of energy and angular momentum along quasi-circular 
orbits. Following a prescription suggested in~\cite{DIS} the badly 
convergent~\cite{Poisson95} PN-Taylor series giving the angular momentum
flux is {\it resummed} by means of Pad\'e approximants.

Our general motivation for studying this problem is: (i) to gain 
information on black hole plunges in a regime ($\mu\ll M$) that is not 
yet accessible to full numerical simulations, and (ii) to learn how to 
match analytical results to numerical ones in a situation where it is 
relatively easy to perform many, controllable high-accuracy numerical 
simulations. This paper will be mainly devoted to the discussion of our 
numerical framework; see~\cite{DN06} for a thorough comparison between 
analytical and numerical results.

\section{Relative dynamics of extreme-mass-ratio binary black holes}
\label{sec:particle}

In the EOB approach the relative dynamics of a binary system of masses $m_1$ 
and $m_2$ is described
by a Hamiltonian $H_{\rm EOB}(M,\mu)$ and a radiation reaction force 
${\cal F}_{\rm EOB}(M,\mu)$, where $M\equiv m_1+m_2$ and $\mu\equiv m_1m_2/M$. 
In the general comparable-mass case $H_{\rm EOB}$
has the structure $H_{\rm EOB}(M,\mu)=M\sqrt{1+2\nu(\hat{H}_\nu - 1)}$
where $\nu\equiv \mu/M\equiv m_1m_2/(m_1+m_2)^2$ is the symmetric mass ratio.
In the test mass limit that we are considering, $\nu\ll 1$, we can expand
$H_{\rm EOB}$ in powers of $\nu$. After subtracting inessential constants
we get a Hamiltonian per unit ($\mu$) mass 
$\hat{H}=\lim_{\nu \to 0}(H-{\rm const.})/\mu=\lim_{\nu\to 0}\hat{H}_\nu$ 
of the form
\begin{equation}
\label{schw:ham}
\hat{H} = \sqrt{ A(\hat{r}) \left(1+\frac{p_r^2}{B(\hat{r})}+\frac{p_{\varphi}^2}{\hat{r}^2} \right) } \ .
\end{equation}
Here we have introduced the dimensionless variables $\hat{r}\equiv r/M$, 
$p_r\equiv\hat{P}_r\equiv P_r/\mu$ and $p_{\varphi}\equiv\hat{P}_\varphi/M\equiv P_\varphi/(\mu M)$.
The functions $A(\hat{r},\nu)$ and $B(\hat{r},\nu)$ entering the Hamiltonian $\hat{H}_\nu$ 
are metric coefficients entering the effective one body metric  
\begin{equation}
\rmd s^2 = g_{\mu\nu}\rmd x^\mu \rmd x^\nu = -A(\hat{r},\nu) \rmd t^2 + B(\hat{r},\nu) \rmd r^2 
+ r^2\left(\rmd\theta^2+\sin^2\theta \rmd\varphi^2\right) \ .
\end{equation}
In the limit $\nu\to 0$ in which we shall use~(\ref{schw:ham}) the effective 
metric functions $A(\hat{r},\nu)$ and $B(\hat{r},\nu)$ reduce to the well-known 
Schwarzschild expressions $A(\hat{r},0)= (B(\hat{r},0))^{-1} = 1-2/\hat{r}$. Since 
the radiation generation process will be studied in terms of the Regge-Wheeler 
tortoise coordinate $r_*=r+2M\log(r/(2M)-1)$, it is useful to have the Hamiltonian 
explicitly written in terms of $r_*$ and its conjugate 
momentum $p_{r_*}$.  This amounts to performing a canonical transformation 
$(\hat{r},p_r)\rightarrow (\hat{r}_*,p_{r_*})$. The invariance of the action,
$p_{r_*}\rmd\hat{r}_*=p_r \rmd\hat{r}$ yields $p_r=(\rmd\hat{r}_*/\rmd\hat{r})p_{r_*}= A^{-1}p_{r_*}$, so that the new 
Hamiltonian function becomes
\begin{equation}
\hat{H}(\hat{r}_*,p_{r_*}) = \sqrt{ A\left( 1+ \frac{p_{\varphi}^2}{\hat{r}^2} \right)+p_{r_*}^2} \ .
\end{equation}
Hamilton's canonical equations in the equatorial plane ($\theta=\pi/2$)
yield
\begin{eqnarray}
\label{eq_rdot_star}
\dot{\hat{r}}_*  & = &  \frac{p_{r_*}}{\hat{H}}  \ , \\
\dot{\hat{r}}    & = &  \frac{A}{\hat{H}}p_{r_*}\equiv v_r \ , \\ 
\dot{\varphi}    & = & \frac{A}{{\hat H}}\frac{p_{\varphi}}{\hat{r}^2}\equiv \omega \ , \\
\dot{p}_{r_*}    & = & -\frac{\hat{r}-2}{\hat{r}^3\hat{H}}\left[p_{\varphi}^2\left(\frac{3}{\hat{r}^2}-\frac{1}{\hat{r}}\right)+1\right] \ ,\\
\label{eq:angmom_diss}
\dot{p}_{\varphi}& = &  \hat{\cal F}_{\varphi} \equiv 
-\frac{32}{5}\nu \omega^5 \hat{r}^4 \frac{\hat{f}_{\rm DIS}(\omega\hat{r})}{1-\sqrt{3}\omega\hat{r}} \ .
\end{eqnarray}
On the r.h.s. of the last equation we have introduced the second ingredient
of the EOB approach: the radiation reaction force $\hat{\cal F}_{\varphi}$.
The function $\hat{f}_{\rm DIS}(\omega\hat{r})$ is the ``factored flux function'' 
of~\cite{DIS} scaled to the Newtonian (quadrupole) flux. Reference~\cite{DIS} 
has shown that the sequence of near-diagonal Pad\'e approximants of 
$\hat{f}_{\rm DIS}(\omega\hat{r})$  exhibits, during the quasi-circular 
inspiral phase ($\hat{r}>6$), a good convergence toward the 
exact result known numerically~\cite{Poisson95} in the $\nu\to0$ limit. 
Here we shall use the $\nu\to 0$ limit of the 2.5~PN accurate 
$\hat{f}_{\rm DIS}(\omega\hat{r})$ ({\it i.e.}, equations~(3.28)-(3.36) 
of \cite{BD00} in the $\nu=0$ limit). Following the suggestion 
of~\cite{DG06}, we expressed, in~(\ref{eq:angmom_diss}), $\hat{\cal F}_\varphi$
as a function of the two variables $\omega$ and $\hat{r}$ with the hope
that this expression remains approximately valid during the plunge ($\hat{r}<6$).
We shall give evidence below that this hope is fulfilled. Note also that 
the flux used in constructing $\hat{\cal F}_{\varphi}$,
in~(\ref{eq:angmom_diss}), refers to the momentum flux {\it at infinity}.
Reference~\cite{Poisson:1994yf} has shown that the flux down the event 
horizon (of the background Schwarzschild black hole of mass $M$) would 
correspond to a (here negligible) 4~PN effect.

\subsection{Quasi-circular orbits and transition from inspiral to plunge}
\label{sbsc:part_dyn}

\begin{figure}[t]
\includegraphics[width=5.50 cm]{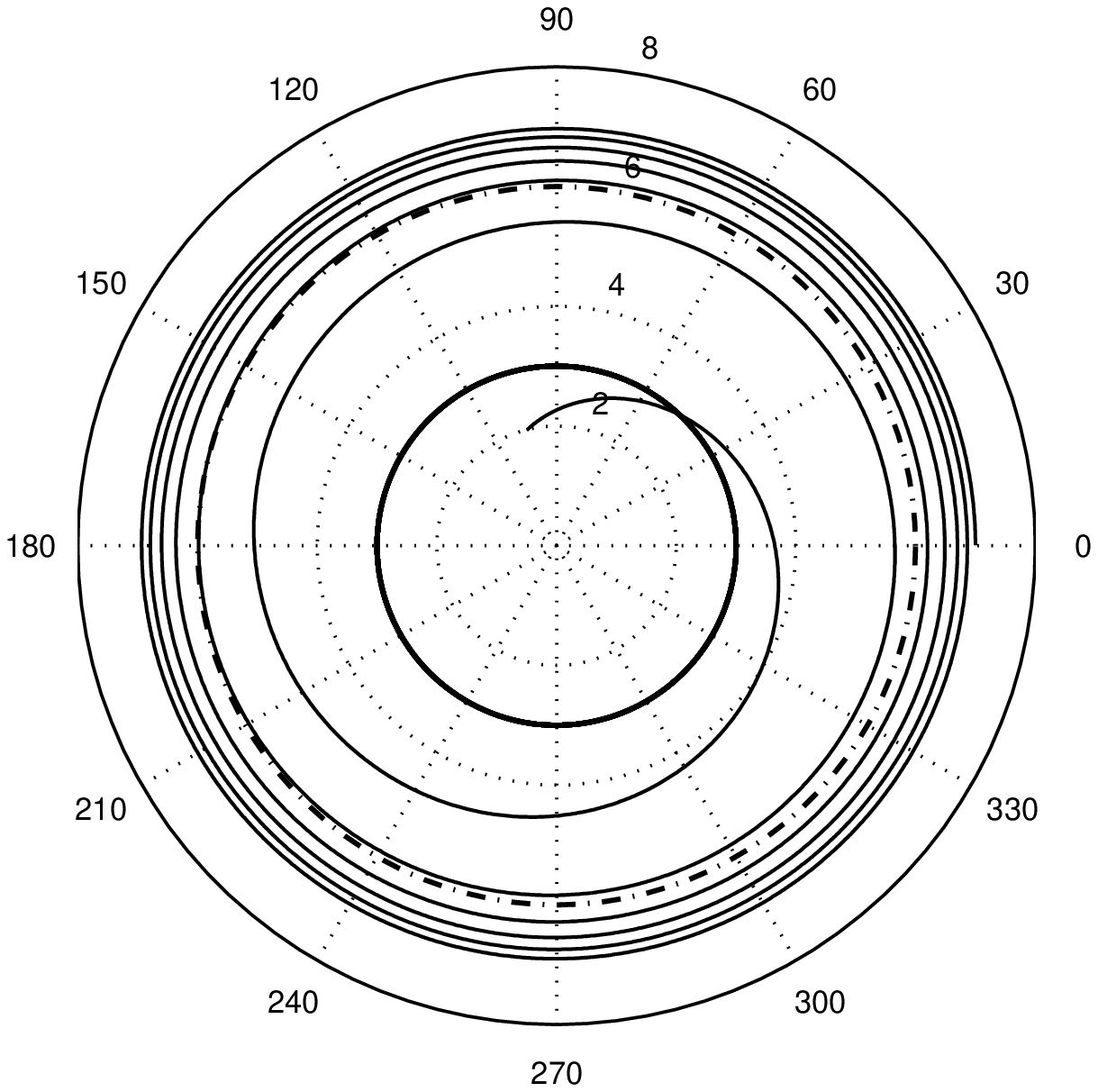}\hspace{0.5 cm}
\includegraphics[width=6.50 cm]{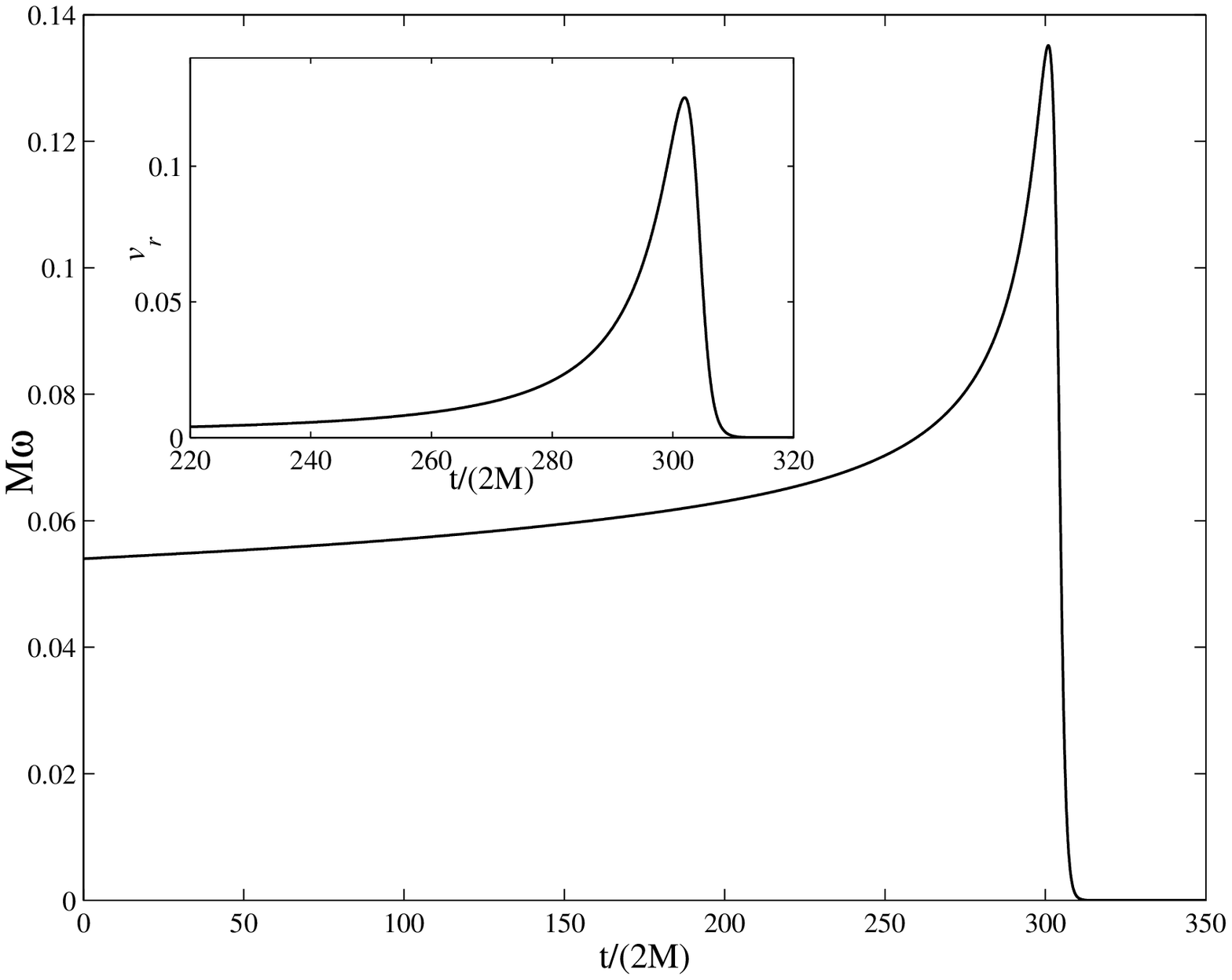}
\fl \caption{\label{label:fig_1}{\it Left panel}: Plunge relative orbit from 
$r=7M$. The circle represented with a thick dashed line is the LSO at $r=6M$. 
The thick solid line circle represents the light ring at $r=3M$. {\it Right panel}: 
Radial velocity $v_r=\dot{\hat{r}}$ and orbital angular frequency $\omega=\dot{\varphi}$ 
versus coordinate time.}
\end{figure}

Following Section ~IV(A) of~\cite{BD00} and taking the $\nu\to 0$ limit
in all terms except the crucial $\hat{{\cal F}}_\varphi={\cal O}(\nu)$,
we start the orbit at some given initial radius  $\hat{r}$ and initial 
phase $\varphi$ with corresponding momenta:
\begin{eqnarray}
p_{\varphi} &\equiv& j_{\rm adiab}= \frac{\hat{r}}{\sqrt{\hat{r} - 3}} \ , \\
p_r         &=& -\frac{1}{C_r}\frac{B_r}{\omega_r^2} \ ,
\end{eqnarray} 
where we explicitly have (with $\hat{H}_0\equiv \hat{H}(p_r=0)$)
\begin{eqnarray}
C_r        &=&  \frac{A^2(\r)}{\hat{H}_0 } \ ,  \\
\omega_r^2 &=&  \frac{1}{(\hat{H}_0)^2}A^2(\r)\frac{\r-6}{\r^3(\r-3)} \ , \\
B_r        &=& -\frac{2 j_{\rm adiab}}{(\hat{H}_0)^2}A^2(\r)\frac{\r-3}{\r^4}\hat{{\cal F}}_\varphi \ , 
\end{eqnarray}
Figure~\ref{label:fig_1} shows the 
trajectory determined by this kind of initial $\hat{r}=7$, $\varphi=0$ and 
$\nu=0.01$ (for the sake of convenience we fix $M=1$ in the numerical calculations); 
after $\sim 5$ orbits the trajectory crosses the Last Stable Orbit (LSO) at 
$\hat{r}=6$ and then ``plunges'' (in a quasi-circular way $v_r\ll 1$, see inset
in the right panel) into the black hole. It has to be noted that the orbital 
angular frequency $M\omega$ reaches a sharp maximum nearly at the light ring  
crossing ($\hat{r}=3$).
%
%
\section{Gravitational wave generation in the extreme mass ratio limit}
\label{sec:perturbations}

We now complete the description of an extreme-mass-ratio binary system
by computing the gravitational wave emission driven by the system during 
the inspiral and the plunge. 

Let us recall that non-spherical linear metric 
perturbations around a Schwarzschild black hole can be expanded in 
scalar, vector and tensor spherical harmonics that are called even-parity 
[or odd-parity] if they transform under parity as $(-1)^{\ell}$ 
[or $(-1)^{\ell+1}$] and are decoupled because the black hole is 
non rotating. The seven even-parity metric multipoles [or the three 
odd-parity multipoles] can be combined together so that Einstein 
equations yield a wave-like equation for  
the even-parity [or odd-parity] gauge-invariant master variable 
$\Psi^{(\rm e)}_{\ell m}$ [or $\Psi^{(\rm o)}_{\ell m}$] with 
corresponding (Zerilli-Moncrief/Regge-Wheeler) potential 
$V_{\ell}^{(\rm e/o)}$. In Schwarzschild coordinates and 
in the presence of a general matter source driving the 
perturbation of the spacetime, they read
\begin{equation}
\label{eq:ZMRW}
\de^2_t\Psi^{(\rm e/o)}_{\ell m}-\de^2_{r_*}\Psi^{(\rm e/o)}_{\ell m}
+V_{\ell}^{(\rm e/o)}\Psi^{(\rm e/o)}_{\ell m} = S^{(\rm e/o)}_{\ell m} \ .
\end{equation}
The well-known expressions of the potentials and the less well-known 
expressions of the sources for a general stress-energy tensor can be found 
in~\cite{MP05,NR05}.  We recall that from $\Psi^{(\rm e)}_{\ell m}$ and 
$\Psi^{(\rm o)}_{\ell m}$ one can easily obtain, when considering the 
limit $r\to\infty$, the $h_+$ and $h_\times$ gravitational-wave polarization 
amplitude, and the emitted power and angular momentum flux according to
\begin{eqnarray}
\label{eq:GW_amplitude}
\fl h_+-{\rm i}h_{\times}&=&\frac{1}{r}\sum_{\ell\geq 2,m}\sqrt{\frac{(\ell+2)!}{(\ell-2)!}}
   \left(\Psi^{(\rm e)}_{\ell m}
  +{\rm i}\Psi^{(\rm o)}_{\ell m}\right)
  \;_{-2}Y^{\ell m} \ , \\
\label{eq:dEdt}
\fl \dot{E}&=&\frac{1}{16\pi}\sum_{\ell\geq 2,m}\frac{(\ell+2)!}{(\ell-2)!}
	\left(\left|\dot{\Psi}^{(\rm o)}_{\ell m}\right|^2 + 
        \left|\dot{\Psi}^{(\rm e)}_{\ell m}\right|^2\right)\;, \\
\label{eq:dJdt}
\fl \dot{J}&=& \frac{1}{32\pi}\sum_{\ell\geq 2,m}\bigg\{{\rm i}m\frac{(\ell+2)!}{(\ell -2)!}
\left[\dot{\Psi}^{(\rm e)}_{\ell m}\Psi^{(\rm e)*}_{\ell m}
+\dot{\Psi}_{\ell m}^{({\rm o})}\Psi^{(\rm o)*}_{\ell m}\right] + c.c.\bigg\} \ .
\end{eqnarray}
where the overdot stands for time-derivative and 
$\;_{-2}Y^{\ell m}\equiv\,_{-2}Y^{\ell m}(\theta,\varphi)$ are 
the $s=2$ spin-weighted spherical harmonics~\cite{goldberg67}. 
In our physical setting ($\nu\to 0$) the sourcing stress-energy
tensor is, to leading order, that of a test particle following
the (relative) dynamics described above. We can then use the gauge-invariant 
multipoles of the stress-energy tensor of a point particle as 
explicitly given in Appendix A of~\cite{NR05}.
Since the Zerilli-Moncrief and Regge-Wheeler equations are solved on 
a $r_*$ grid, it is natural to express the source in terms of 
$\delta(r_*-R_*(t))$ where $R_*(t)$ denotes the ``particle tortoise
coordinate'', by contrast to a general ``field coordinate'' $r_*$.
In addition, since the particle dynamics is written using canonical 
variables, it is also natural to express the source in terms of the
particle coordinates $R_*$ and $\Phi$ and their (rescaled) conjugate momenta  
$\hat{P}_{r_*}$ and $\hat{P}_{\varphi}$. This has the advantage 
[over the use of the $r$ coordinate and $\delta(r-R(t))$ ] of allowing
one to push the evolution in time as much as one as one wants when
$r\to 2M$,  without introducing spurious boundary effects that may 
spoil the waveforms (and in particular the late-time tail).
The coordinate transformation $r\to r_*$ (with $\rmd r/\rmd r_*=1-2M/r=A(r)$ )
implies the relations
\begin{eqnarray}
\fl \delta(r-R(t))      &=&  A(r)^{-1}\delta(r_*-R_*(t)) \ ,\\
\fl \de_r\delta(r-R(t)) &=& -\frac{2M}{A(r)^2 r^2}\delta(r_*-R_*(t))+\frac{1}{A(r)^2}\de_{r_*}\delta(r_*-R_*(t)) \ .
\end{eqnarray}
Straightforward algebra permits then to derive from the results of~\cite{NR05} the following
explicit expressions for the source terms in~(\ref{eq:ZMRW})
\begin{eqnarray}
\fl S^{(\rm e)}_{\ell m} &=& -\frac{16\pi\mu Y^*_{\ell m}}{r\hat{H}\lambda[(\lambda-2)r+6M]}
\Bigg\{\left(1-\frac{2M}{r}\right)\left(\hat{P}_{\varphi}^2+r^2\right)\de_{r_*}\delta(r_*-R_*(t))\nonumber\\
\fl &+&\Bigg\{-2{\rm i}m\left(1-\frac{2M}{r}\right)\hat{P}_{R_*}\hat{P}_{\varphi} + \left(1-\frac{2M}{r}\right)\Bigg[3M\left(1
+\frac{4\hat{H}^2 r}{(\lambda-2)r+6M}\right)\nonumber\\
\fl &-&\frac{r\lambda}{2}+\frac{\hat{P}_\varphi^2}{r^2(\lambda-2)}\left[r(\lambda-2)(m^2-\lambda-1)+2M(3m^2-\lambda-5)\right]\nonumber\\
\fl &+&\left(\hat{P}^2_{\varphi}+r^2\right)\frac{2M}{r^2}\Bigg]\Bigg\}\delta(r_*-R_*(t))\Bigg\} \ ,
\end{eqnarray}
\begin{eqnarray}
\label{src:odd}
\fl S^{(\rm o)}_{\ell m} &=& \frac{16\pi\mu\de_\theta Y^*_{\ell m}}{r\lambda(\lambda-2)}
\Bigg\{\left[\left(\frac{\hat{P}_{R_*}\hat{P}_{\varphi}}{\hat{H}}\right)_{,t}-2\hat{P}_\varphi\frac{r-2M}{r^2}
-{\rm i}m\frac{r-2M}{r^3}\frac{\hat{P}_{R_*}\hat{P}_{\varphi}^2}{\hat{H}^2}\right]\delta(r_*-R_*(t)) \nonumber\\
\fl &+&\left(1-\frac{P_{R_*}^2}{\hat{H}}\right)\hat{P}_{\varphi}\de_{r_*}\delta(r_*-R_*(t)) \Bigg\} \ ,
\end{eqnarray}
where $\lambda\equiv\ell(\ell +1)$. In the above equations we have chosen to write the sources in the 
functional form
\begin{equation}
\label{source:standard}
\fl S^{(\rm e/o)}_{\ell m} = G^{(\rm e/o)}_{\ell m}(r,t)\delta(r_*-R_*(t)) + F^{(\rm e/o)}_{\ell m}(r,t)\de_{r_*}\delta(r_*-R_*(t)) \ ,
\end{equation}
with $r$-dependent (rather than $R(t)$-dependent) coefficients $G(r)$, $F(r)$.
Note that the time dependence of $F(r,t)$ and $G(r,t)$ comes from the dependence
on $\hat{H}$, $\hat{P}_{R_*}$ and $\hat{P}_{\varphi}$. 
By exploiting the properties of the $\delta$-function, we can also rewrite the 
sources in the functional form
\begin{equation}
\label{source:reload}
\fl S^{(\rm e/o)}_{\ell m} = \tilde{G}^{(\rm e/o)}_{\ell m}(R_*(t)) \delta(r_*-R_*(t)) 
+ F^{(\rm e/o)}_{\ell m}(R_*(t))\de_{r_*}\delta(r_*-R_*(t)) \ ,
\end{equation}
where  
\begin{equation}
\fl \tilde{G}^{(\rm e/o)}_{\ell m}(R_*) = G^{(\rm e/o)}_{\ell m}(R_*) 
-\left . \frac{\rmd F^{(\rm e/o)}_{\ell m}}{\rmd r_*}\right\vert_{r_*=R_*} \ .
\end{equation}
Expressions~(\ref{source:standard}) and~(\ref{source:reload}) are mathematically equivalent,
for a distributional source, but become numerically (slightly) different when the 
$\delta$-function is approximated on the $r_*$ numerical domain by means of a narrow Gaussian. 

\section{Numerical framework, tests and comparison with the literature}
\label{sec:nums}
\subsection{Numerical procedure}
\label{sbsc:waveeqs}
We solve the equations given in section~\ref{sec:particle} for the particle 
dynamics using a standard fourth-order Runge-Kutta algorithm with adaptive 
stepsize~\cite{nr}. Then we insert the resulting position and momenta in
the  source terms $S_{\ell m}^{(\rm e/o)}$ using a Gaussian-function 
representation of $\delta(r_*-R_*(t))$ (see below).
The corresponding Zerilli-Moncrief and Regge-Wheeler equations are then
numerically solved, using standard techniques~\cite{nr,KOG}, in the time 
domain (for each multipole $(\ell,m)$) by discretizing the 
$r_*$ axis ( $r_*\in[r_*^{\rm min},r_*^{\rm max}]$ ) with a uniform grid 
spacing $\Delta r_*$.
In particular, the solution can be computed either by means of a centred
second-order finite differencing algorithm, or by the following implementation 
of the Lax-Wendroff method (that is the one usually preferred). 
In this second case, it is convenient to rewrite the wave-equations as 
a first-order flux conservative system (with sources) in the form
$\de_t {\bf U} + \de_{r_*}{\bf F}= {\bf S}$, 
where we have defined the vector of ``conserved quantities'' ${\bf U}$ 
and the ``fluxes'' ${\bf F}$ as
\begin{equation}
{\bf U}\equiv
\left(\begin{array}{c}
\Psi^{(\rm e/o)}_{\ell m} \cr
w 
\end{array}\right) 
\qquad {\bf F}\equiv 
\left(\begin{array}{c}
\Psi^{(\rm e/o)}_{\ell m} \cr
-w
\end{array}\right)
\end{equation}
where $w = \de_t\Psi^{(\rm e/o)}_{\ell m}+\de_{r_*}\Psi^{(\rm e/o)}_{\ell m}$ 
and the vector  {\bf S} is
\begin{equation}
{\bf S} = \left(\begin{array}{c}
w \cr
V_{\ell}^{(\rm e/o)}\Psi^{(\rm e/o)}_{\ell m}+S_{\ell m}^{(\rm e/o)} \cr
\end{array}\right) \ ,
\end{equation}
Denoting by $j$ the spatial grid-point index, $n$ the time level and $\Delta t$ the time step, 
the explicit time-advancing algorithm reads
\begin{equation}
\fl {\bf U}^{n+1}_j={\bf U}^n_j-\frac{\Delta t}{2\Delta r_*}\left[{\bf F}^n_{j+1}-{\bf F}^n_{j-1}\right] 
               +\frac{\Delta t^2}{2\Delta r_*^2}\left[{\bf F}^n_{j+1}-2{\bf F}^n_j+{\bf F}^n_{j-1}\right]\mathbb{A}+\Delta t{\bf S}^n_j \ ,
\end{equation}
where we introduced the matrix $\mathbb{A}\equiv {\rm diag}(1,-1)$. This algorithm
is stable under the standard Courant-Friedrichs-Levy~\cite{KOG} condition 
$\Delta t/\Delta r_*<1$. We typically use $\Delta t/\Delta r_*=0.9$.
On the boundaries we impose ingoing (at $r_*^{\rm min}$) and outgoing (at $r_*^{\rm max}$) 
standard Sommerfeld conditions. This efficiently suppress reflections from
the boundaries apart from tiny effects that can modify (if we do not use sufficiently
large grids) the power-law tail at the end of the black hole ringdown phase. 
[Improved non-reflecting boundary conditions for Zerilli-Moncrief and Regge-Wheeler 
equations have been recently discussed~\cite{lau05}].

\subsection{Approximating the $\delta$-function}
\label{sbsc:delta}
We approximate the $\delta$-function that appear in the source terms by a smooth
function $\delta_\sigma(r_*)$ (and $\de_{r_*}\delta(r_*)$ by $\de_{r_*}\delta_\sigma(r_*)$).
We use\footnote{Note that this smoothing of the distributional stress-energy
tensor into a formally extended one only concerns the calculation of the
waveform. The dynamics, equations~(\ref{eq_rdot_star})-(\ref{eq:angmom_diss}),
is that of a $\delta$-function stress-energy tensor. We use this smoothing only 
as a numerical technique, and we have checked that we were in the convergent regime.}
\begin{equation}
\label{eq:delta}
\delta(r_*-R_*(t))\rightarrow \;\frac{1}{\sigma\sqrt{2\pi}}\exp \left[{-\frac{(r_*-R_*(t))^2}{2\sigma^2}}\right] \ ,
\end{equation}
with $\sigma\geq\Delta r_*$. In practice $\sigma\simeq\Delta r_*$ works well thanks to
the effective averaging entailed by the fact that $R_*(t)$ is not restricted to the $r_*$
grid, but varies nearly continuously on the $r_*$ axis. 
In the continuum limit ($\sigma\simeq\Delta r_*\to 0$) the derivatives 
of both $\Psi^{(\rm e)}_{\ell m}$ and $\Psi^{(\rm o)}_{\ell m}$ would be discontinuous at the location of 
the particle $r_*=R_*$, generating numerical noise (Gibbs phenomenon) if standard numerical 
methods are used. However, the smoothing of $\delta(r_*-R_*(t))$, together with the use of 
a numerical method like Lax-Wendroff (that has a bit of numerical dissipation built in) 
avoids any problems related to high frequency oscillations and provides us with a clean 
and stable evolution~\footnote{Note that the use of a method without dissipation, 
like a standard leapfrog, introduces high frequency noise if $\sigma$ is too small. 
For this reason we prefer to use the Lax-Wendroff technique for most of the simulations.}.
The important scale in our problem is $M$ which determines the ``width'' of the Zerilli 
potential. We found that a resolution of $\sigma\simeq\Delta r_*=0.01M$ is 
small enough to ensure numerical convergence to the continuum solution.

\subsection{Numerical tests: circular orbits and radial plunge}
\label{sbsc:circular}
We have tested the reliability of our code by recomputing results for simple
geodesic trajectories ($\hat{{\cal F}}_\varphi=0$) that have been already 
obtained in the literature using both time-domain and frequency domain 
approaches, as well as different treatments for the particle and different 
expressions for the sources. 

First of all we discuss the case of circular orbits.
Table~\ref{table_one} lists the energy and angular momentum fluxes  
(at $r_{\rm obs}=1000M$) up to the $\ell=4$ multipole for an orbit of radius 
$r=7.9456$. This permits a direct comparison with the results of 
Martel~\cite{martel04} (see also~\cite{Sopuerta_2005gz}), 
that we include in the table as well for the sake of 
completeness. For this test run we consider a (relatively) coarse
resolution of $\Delta r_*=0.02M$; this is enough to have a very small
difference with Martel results for the $\ell=2$ multipoles, but the
agreement obviously worsens for higher multipoles, for which it is 
necessary to increase the resolution. That is the reason why, when 
discussing the real plunge phenomenon driven by radiation reaction 
we shall use resolutions up to $\Delta r_*=5\times 10^{-3}$ to 
properly capture the behaviour of the highest multipoles.

Our second test concerns the waveform of a particle
plunging radially into the black hole from a finite distance 
(where it starts at rest). This problem has received some attention
in the recent literature~\cite{LP97,MP02}, thereby extending 
(by means of both frequency-domain and time domain approaches) 
the pioneering analysis of Davis {\it et al.}~\cite{DRPP,DRT} 
of the early 70s for a particle plunging from an infinite distance.

%
%
\begin{table}
\caption{\label{table_one}Energy and angular momentum fluxes
extracted at $r_{\rm obs}=1000M$ for a particle orbiting the
black hole on a circular orbit of radius $r=7.9456$. Comparison 
with the results of Martel~\cite{martel04}.} 
\begin{indented}
\lineup
\item[]\begin{tabular}{@{}*{8}{l}}
\br  
 $\ell$ & $m$ &    $(\dot{E}/\mu^2)_{\rm here}$      & ($\dot{E}/\mu^2)_{\rm Martel}$  & rel. diff. &($\dot{J}/\mu^2)_{\rm here}$ & $(\dot{J}/\mu^2)_{\rm Martel}$ & rel. diff. \cr                        
\mr
    2       &  1  & $8.1998\times10^{-7}$   &  $8.1623\times10^{-7}$   & $0.4\%$ &$1.8365\times10^{-5}$   &  $1.8270\times10^{-5}$  &  $0.5\%$\cr
            &  2  & $1.7177\times10^{-4}$   &  $1.7051\times10^{-4}$   & $0.7\%$ &$3.8471\times10^{-3}$   &  $3.8164\times10^{-3}$  &  $0.5\%$\cr
    3       &  1  & $2.1880\times10^{-9}$   &  $2.1741\times10^{-9}$   & $0.6\%$ &$4.9022\times10^{-4}$   &  $4.8684\times10^{-8}$  &  $0.7\%$\cr
            &  2  & $2.5439\times10^{-7}$   &  $2.5164\times10^{-7}$   & $1.1\%$ &$5.6977\times10^{-6}$   &  $5.6262\times10^{-6}$  &  $1.2\%$\cr
            &  3  & $2.5827\times10^{-5}$   &  $2.5432\times10^{-5}$   & $1.5\%$ &$5.7846\times10^{-4}$   &  $5.6878\times10^{-4}$  &  $1.7\%$\cr
    4       &  1  & $8.4830\times10^{-13}$  &  $8.3507\times10^{-13}$  & $1.6\%$ &$1.8999\times10^{-11}$  &  $1.8692\times10^{-11}$ &  $1.6\%$\cr
            &  2  & $2.5405\times10^{-9}$   &  $2.4986\times10^{-9}$   & $1.7\%$ &$5.6901\times10^{-8}$   &  $5.5926\times10^{-8}$  &  $1.7\%$\cr
            &  3  & $5.8786\times10^{-8}$   &  $5.7464\times10^{-8}$   & $2.3\%$ &$1.3166\times10^{-6}$   &  $1.2933\times10^{-6}$  &  $1.8\%$\cr
            &  4  & $4.8394\times10^{-6}$   &  $4.7080\times10^{-6}$   & $2.7\%$ &$1.0838\times10^{-4}$   &  $1.0518\times10^{-4}$  &  $3.0\%$\cr
\br
\end{tabular}
\end{indented}
\end{table}
\begin{figure}[t]
\begin{center}
\includegraphics[width=7.50 cm]{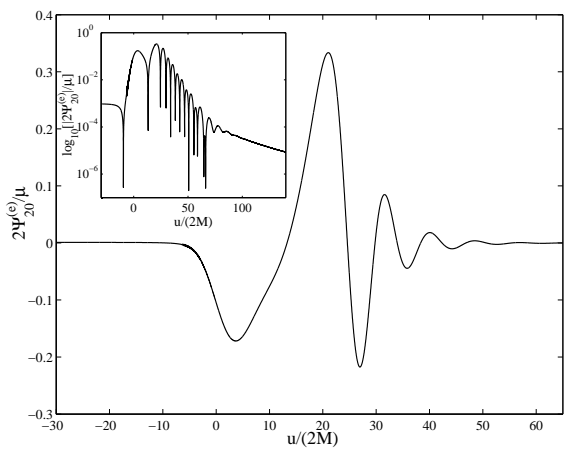}
\fl \caption{\label{fig3}Test of the code: waveform 
(on a logarithmic scale in the inset) for a particle 
plunging radially on the black hole along the 
$z$-axis from $r=10M$}
\end{center}
\end{figure}
We selected the same kind of initial data as described
in references~\cite{LP97,MP02}, so as to have an initial profile of $\Psi^{(\rm e)}_{\ell 0}$
that is conformally flat. However, since we have slightly ``smeared'' 
the $\delta$-function, we cannot use the ``discontinuous'' analytical initial 
data of references~\cite{LP97,MP02}. We numerically solve the Hamiltonian 
constraint by writing it as a tridiagonal system that is then 
solved using a standard $LU$ decomposition. Figure~$\ref{fig3}$ displays
(for the case $\ell=2$) the waveform generated by a particle plunging into the 
black hole along the $z$-axis, starting from rest at $r=10M$. 
It has been extracted at $r_{\rm obs}=500M$ and is shown versus the retarded 
time $u=t-r_*^{\rm obs}$. The master function $\Psi^{(\rm e)}_{20}$ has been 
multiplied by a factor two (linked to a different choice of 
normalization in~\cite{LP97,MP02} ) to facilitate the (very satisfactory) 
comparison with the top-right panel of figure~4 in~\cite{MP02} or the top-left panel of figure~6 in~\cite{LP97}.
We used a resolution of $\Delta r_*=0.01M$ with $r_*\in[-800M,1800M]$. 
This avoids the influence of the boundaries and allows one to capture 
the late-time non-oscillatory decay after the QNM ringdown phase 
(see the inset in figure~$\ref{fig3}$).

%
\begin{figure}[t]
\includegraphics[width=6.56 cm]{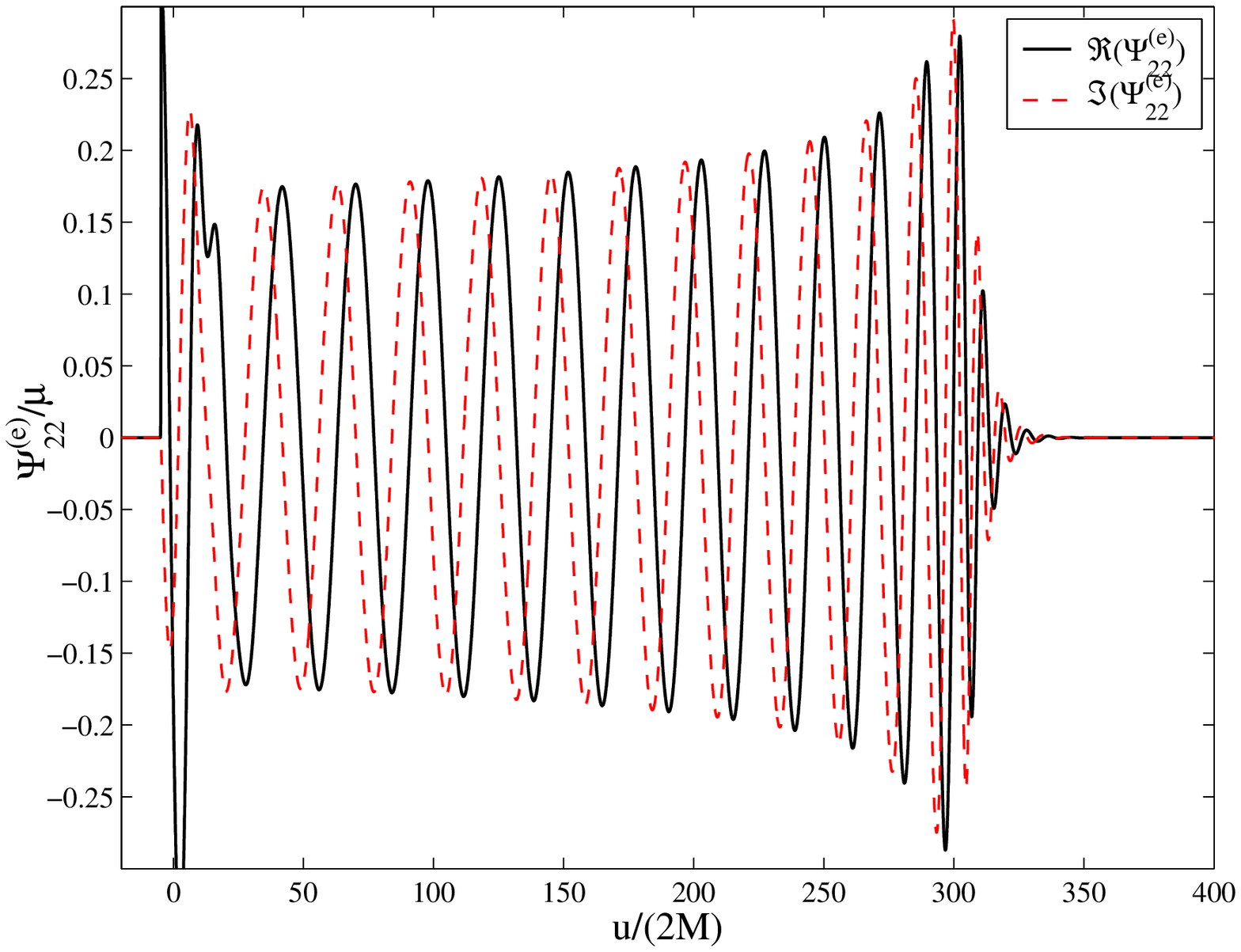}
\includegraphics[width=6.56 cm]{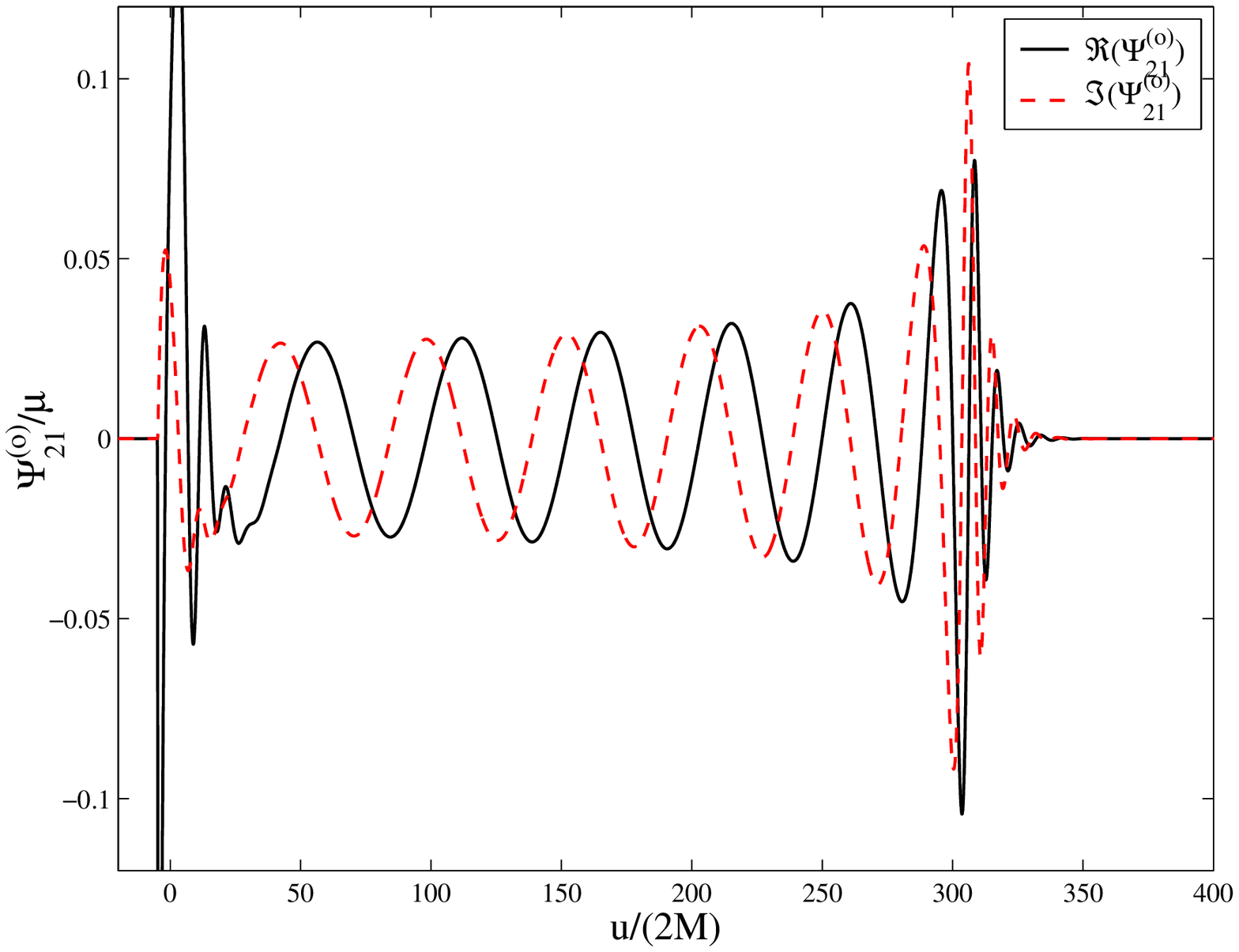}\\
\includegraphics[width=6.56 cm]{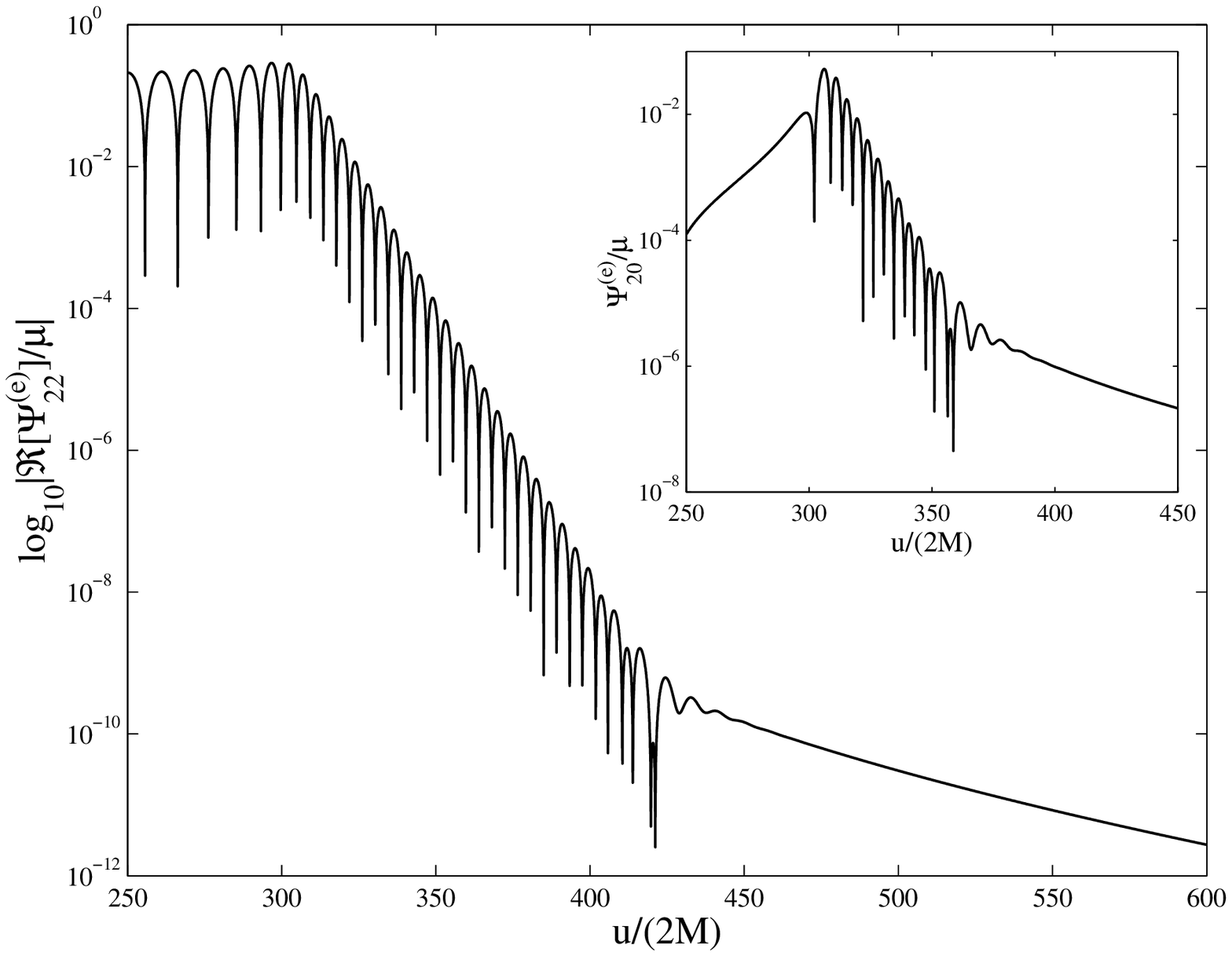}
\includegraphics[width=6.56 cm]{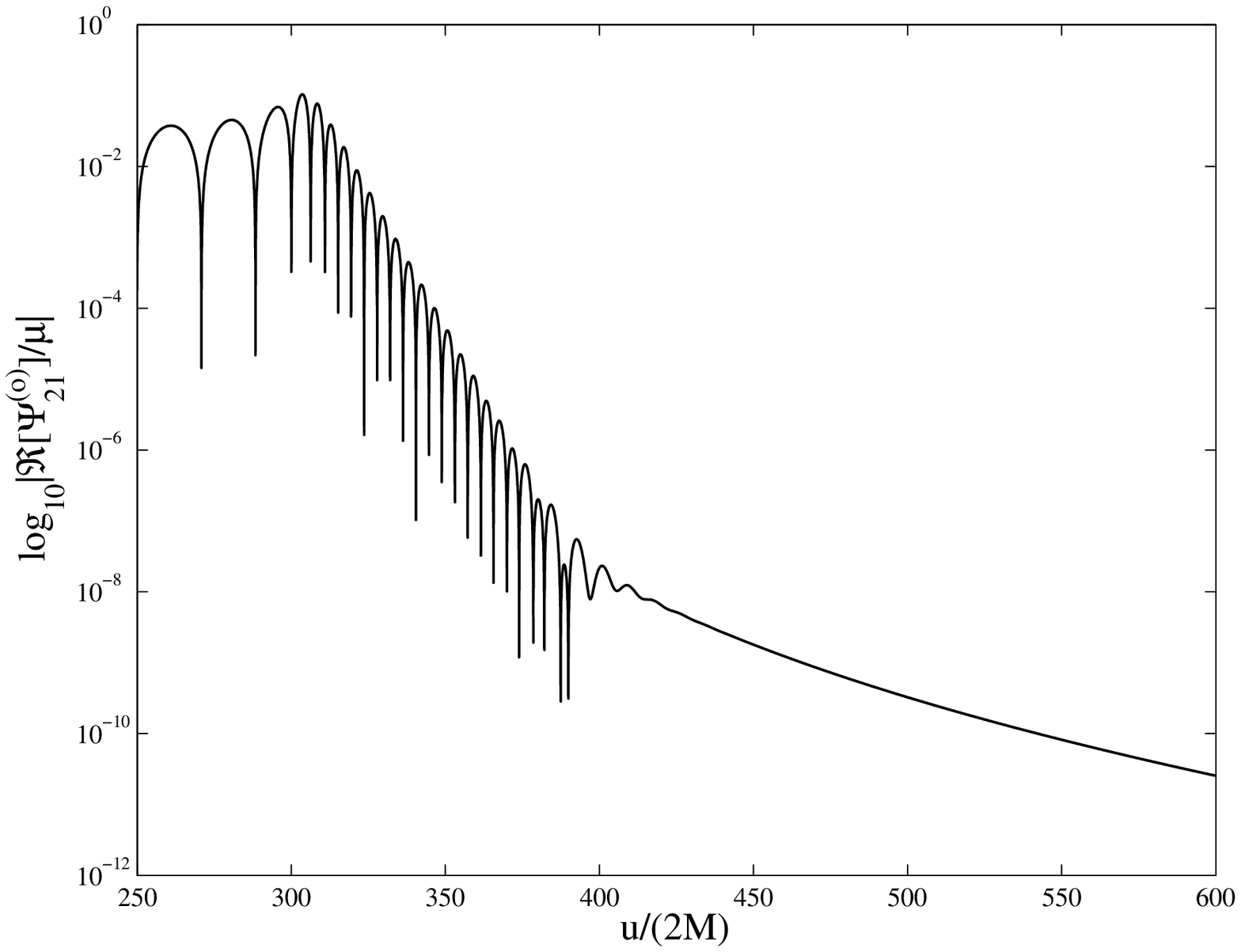}
\fl \caption{\label{fig:waves}Dominant $\ell=m=2$ 
even-parity {\it Left panel} and  and $\ell=2$, $m=1$ odd-parity
{\it Right panel} gravitational wave multipoles  generated by the 
quasi-circular plunge of a particle with $\nu=0.01$ initially at 
$r=7M$.}
\end{figure}

\section{Transition from inspiral to plunge: waveforms}
\label{sec:waveforms}

Now that we have shown that our code can reproduce with good 
accuracy known results for geodesic orbits, let us break new
ground by considering the waveform emitted during the transition 
from inspiral to plunge. We refer here to the dynamics described in 
section~\ref{sbsc:part_dyn}; {\it i.e.}, the particle with $\nu=0.01$ 
is initially at $r=7M$  with dynamical initial data defined by the 
quasi-circular (adiabatic) approximation described there. We also
need initial data for the field perturbations: $\Psi^{(\rm e/o)}_{\ell m}$,
$\de_t\Psi^{(\rm e/o)}_{\ell m}$. To do so, one should in principle select
the solution of the perturbed Hamiltonian and momentum constraints 
which corresponds to the physically desired ``no-incoming radiation
condition''. However, we took a more pragmatic approach. 
As it is often done in the literature, we impose 
$\Psi^{(\rm e/o)}_{\ell m}=\de_t\Psi^{(\rm e/o)}_{\ell m}=0$ at $t=0$. 
This ``bad'' choice of initial data produces an initial burst of 
unphysical radiation, but after a while the gravitational waveform 
is driven by the motion of the source so that it is sufficient not 
to include the early part of the waveform in the final analysis.
Figure~\ref{fig:waves} depicts the dominant $\ell=2$ waveforms 
($m=2$ and $m=1$ multipoles) extracted at $r_{\rm obs}=250M$. For the sake 
of comparison we normalize the waveforms to the mass $\mu$ (see below a 
discussion of the approximate universality of this scaled waveform).
The numerical grid we adopted for this simulation was $r_*\in [-1200M,1200M]$ 
with resolution $\Delta r_*=0.01M$. As we were mentioning above, after an 
initial unphysical burst of radiation the gravitational waveforms are driven 
by the motion of the particle during the quasi-circular inspiral and plunge. 
From the inspection of the particle dynamics we described above, we know it 
crosses the LSO at $u/(2M)\simeq 240$ and passes through the light ring (where
the gravitational perturbation driven by the source is filtered 
by the peak of the potential) at $u/(2M)\simeq 300$ (where both the orbital 
frequency $\omega$ and the radial velocity $v_r$ have a maximum). 
Both $\Psi^{(\rm e)}_{22}$ and $\Psi^{(\rm o)}_{21}$ show a progressive 
increase in frequency and amplitude until $u/(2M)\simeq 300$, where 
a maximum of amplitude is reached. Then follows a QNM dominated 
ringdown phase.

To validate our results, we study the convergence of the waveforms
by considering resolutions $\Delta r_*=(0.02,0.01,0.005)$. For the dominant 
$\ell=m=2$ waveforms we found a convergence rate $\beta\simeq 1.6$, defined
from $\Delta\Psi \propto \Delta r_*^\beta$. Here $\Delta\Psi$ is defined as
the discretized root mean square between the waveform at the resolution 
$\Delta r_*$ and the highest resolution one. 
Generally speaking, a resolution $\Delta r_*=0.01$ is sufficient to
determine the energy and angular momentum radiated with an accuracy that
is of the order of $1\%$ (or better) for the quadrupole modes. Having the
same accuracy for higher modes needs to increase the resolution; however,
we have verified that this is sufficient for having an accuracy (for the
energy) that is not worse  than $6\%$ for the other multipoles up to $\ell=4$ 
(the most sensitive one being $\ell=4$, $m=0$). 

To give some numbers we computed the total energy and angular momentum loss 
during what might (approximately) be called the plunge phase (that is, after
the crossing of the LSO at $6M$). More precisely, we selected 
the part of the waveform for $u/(2M)\geq 240$ which corresponds to 
radii $\hat{r}\leq 5.9865$. Table~\ref{table_two} lists the partial multipolar 
energy and angular momentum losses up to $\ell=4$. These values were computed 
with resolution $\Delta r_*=0.01$ so that the accuracies are of the order 
of $1\%$ for $\ell=2$, $2-4\%$ for $\ell=3$ and of $5-6\%$ for $\ell=4$. 
When one sums all the multipoles, the total energy emitted is found to be 
$M\Delta E/\mu^2\simeq0.51$ and $\Delta J/\mu^2\simeq 4.3$. Only for the sake 
of comparison, let us mention that this ``plunge'' $M\Delta E/\mu^2$ is about 
50 times larger than the ``radial'' (Davis {\it et al.} \cite{DRPP,DRT}) 
one (summed up to $\ell=4$) which amounts to $M\Delta E/\mu^2 = 0.0104$ 
(for a plunge from infinity).

\begin{figure}[t]
\includegraphics[width=6.56 cm]{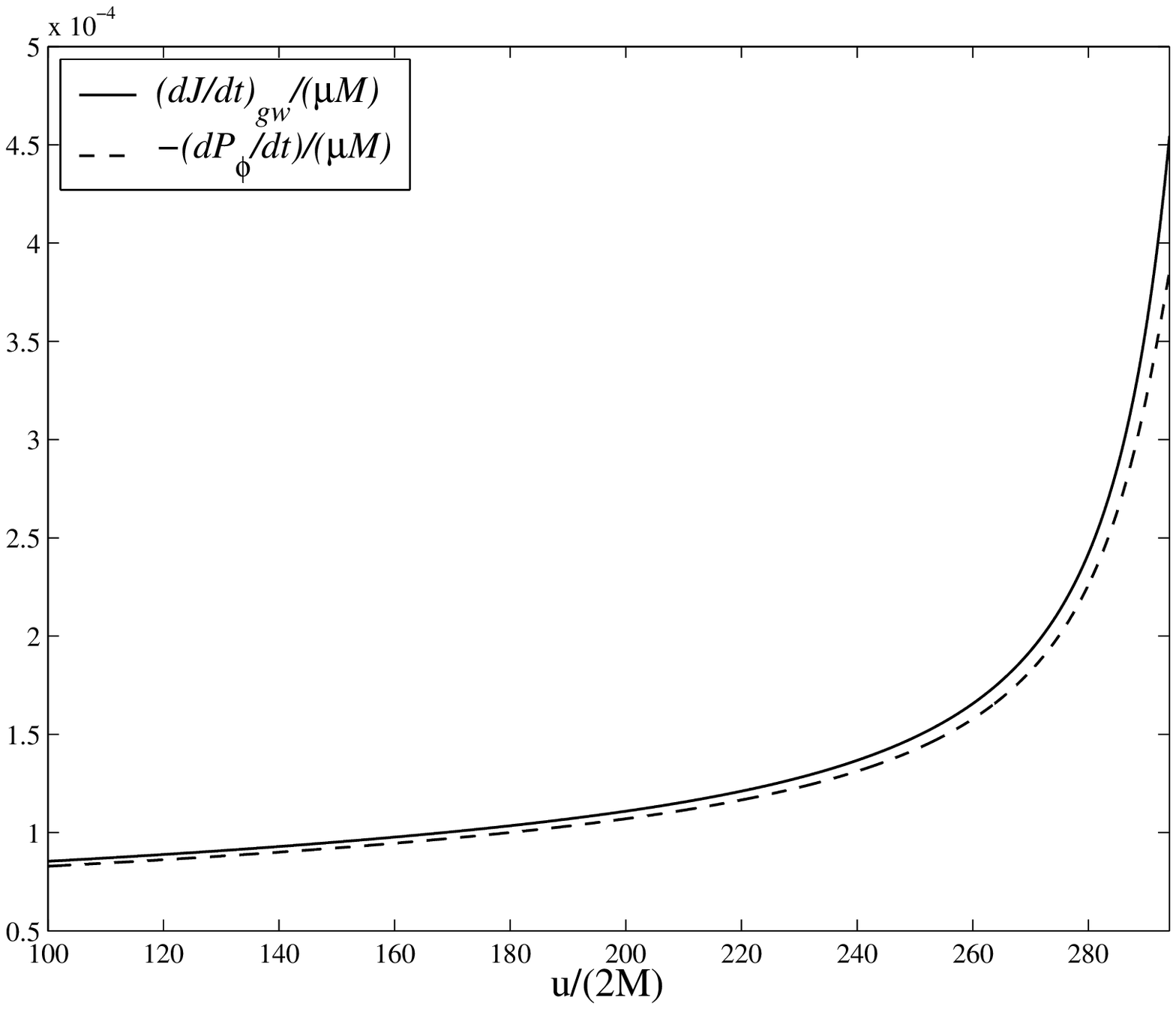}
\includegraphics[width=6.56 cm, height=5.5 cm]{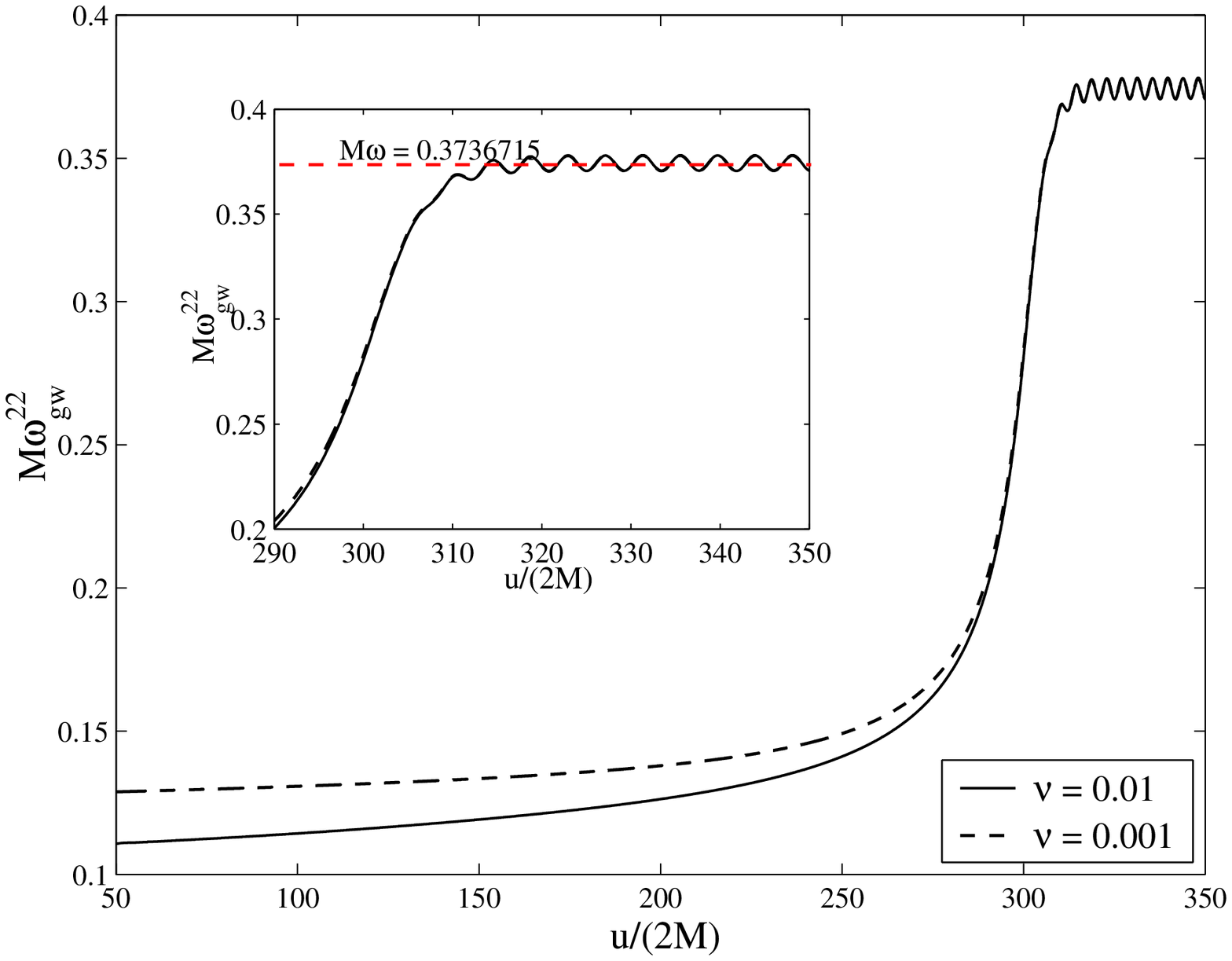}
\fl \caption{\label{fig:versusnu}{\it Left panel}: angular momentum
flux computed from the gravitational waveform compared to the angular
momentum loss assumed in the dynamics with $\nu=0.01$.
{\it Right panel}: Instantaneous gravitational wave frequency for two 
values of $\nu$. The signal is `` approximately universal'' after 
$u/(2M)\simeq 280$, which roughly corresponds to $r\lesssim 5.15$
(``quasi-geodesic plunge''). After $u/(2M)\simeq 300$ starts the 
QNM phase.}
\end{figure}

\begin{table}
%
%
\caption{\label{table_two}Total energy and angular momentum 
extracted at $r_{\rm obs}=250M$ for the ``plunge phase'' ($r<5.9865M$) 
of a dynamics with $\nu=0.01$.} 
\begin{indented}
\lineup
\item[]\begin{tabular}{@{}*{6}{l}}
\br  
 $\ell$ &   $m$  &    &$M\Delta E/\mu^2$ &      & $\Delta J/\mu^2$     \cr                        
\mr
    2   &  $0$   &   & $9.8\times 10^{-4}$ &  & $0$                 \cr
        &  $1$   &   & $2.06\times 10^{-2}$&  & $0.084$             \cr
        &  $2$   &   & $3.31\times 10^{-1}$ &  & $2.994$             \cr
    3   &  $0$   &  & $3.4\times 10^{-5}$  &  & $0$                 \cr
        &  $1$   &  & $5.6\times 10^{-4}$  &  & $1.2\times 10^{-3}$ \cr
        &  $2$   &  & $8.1\times 10^{-3}$  &  & $3.9\times 10^{-2}$ \cr
        &  $3$   &  & $1.05\times 10^{-1}$ &  & $8.5\times 10^{-1}$ \cr
     4  &  $0$   &  & $1.7\times 10^{-6}$  &  & $0$                 \cr
        &  $1$   &  & $2.4\times 10^{-5}$  &  & $3.6\times 10^{-5}$ \cr
        &  $2$   &  & $3.3\times 10^{-4}$  &  & $1.1\times 10^{-3}$ \cr
        &  $3$   &  & $3.5\times 10^{-3}$  &  & $1.8\times 10^{-2}$ \cr
        &  $4$   &  & $4.2\times 10^{-2}$  &  & $3.2\times 10^{-1}$ \cr 
\br
\end{tabular}
\end{indented}
\end{table}
An important consistency check of our approach ( which assumes the specific
PN-resummed radiation reaction force~(\ref{eq:angmom_diss}) )
is to compare the angular momentum loss assumed in the dynamics, 
{\it i.e} $-(\rmd P_{\varphi}/\rmd t )/(\mu M)=-{\hat{\cal F}}_{\varphi}$, to the 
angular momentum flux radiated by the multipolar gravitational waves
$\Psi^{(\rm e/o)}_{\ell m}$ at infinity (up to $\ell=4$).
The left panel of figure~\ref{fig:versusnu} displays $-(\rmd P_{\varphi}/\rmd t )/(\mu M)$ 
(dashed line) versus $(dJ/dt)_{\rm gw}/(\mu M)$ (solid line) up to roughly the
crossing of the light ring. 
The very good agreement between the two curves is a confirmation of the
good convergence of the Pad\'e-resummed radiation reaction force 
$\hat{\cal F}_{\varphi}$, equation~(\ref{eq:angmom_diss}), toward the exact
result. This confirmation goes beyond the tests of~\cite{DIS} which 
were limited to the inspiral phase $\hat{r}>6$. The retarded times
$240\lesssim u/(2M)\lesssim 290$ in figure~\ref{fig:versusnu} correspond
to the plunge phase [the merger (matching) occurring at $u/(2M)\sim 300$].

The question that one can ask at this point is how much the numbers
of table~\ref{table_two} are universal. 
Let us first recall that~\cite{BD00} has shown that the transition between
inspiral and plunge was taking place in a radial domain around the LSO
which scaled with $\nu$ as $r-6M\sim\pm1.89M\nu^{2/5}$, with radial velocity
scaling as $v_r \sim -0.072\nu^{3/5}$.
The specific power $2/5$-th appearing in the radial scaling yields, in the 
numerical case $\nu=0.01$ that we consider in most of this paper, 
$r-6M\sim\pm0.30M$. This means that, formally, when $\nu=0.01$, it is only 
when $6M-r\gg 0.3M$ that we can expect the plunge dynamics to become universal, 
in the sense that it will approach the geodesic which started from the LSO in 
the infinite past with zero radial velocity. If we wanted to reach this universal 
behaviour just below the LSO (say, for $r\leq 5.98$) we would need to use much 
smaller values of $\nu$ (say $\nu\lesssim 10^{-5}$). For larger values of $\nu$ 
we can only hope to see an approximate convergence to universal behaviour near 
the end of the plunge. This is illustrated in the right panel of 
Figure~\ref{fig:versusnu}.
This figure plots, versus retarded time $u$, the instantaneous gravitational 
wave frequency  $M\omega_{\rm gw}^{22}$, where ($\Im\equiv$ imaginary part)
\begin{equation}
\label{inst:wgw}
M\omega_{\rm gw}^{\ell m} = -\Im\left(\frac{\dot{\Psi}_{\ell m}^{(\rm e/o) } }{\Psi_{\ell m}^{(\rm e/o)} }\right) \ 
\end{equation} 
for $\nu=0.01$ (solid line) and $\nu=0.001$ (dashed line). 
At early times, that is when the particle moves along a quasi-circular 
orbit, one has $M\omega_{\rm gw}^{\ell m} \simeq mM\omega=2M\omega$; 
i.e., the double of the orbital frequency. Then one has the transition 
regime mentioned above around the crossing of the LSO, {\it i.e.}, for 
$u/(2M)\simeq 240$. The quasi-universal, quasi-geodesic plunge mentioned 
above starts afterwards around $u/(2M)\simeq 280$, {\it i.e.}, 
$r\simeq 5.15M$ (for $\nu=0.01$)

Then, after crossing the light ring, one enters 
another universal phase: the QNM one. In this QNM phase the gravitational 
wave angular frequency saturates and oscillates around the value 
$M\omega_{\rm gw}^{22}=0.3736715$ (highlighted in the inset). The fine 
structure of this gravitational wave frequency plot will be further 
discussed in~\cite{DN06}.

\begin{figure}[t]
\includegraphics[width=6.56 cm]{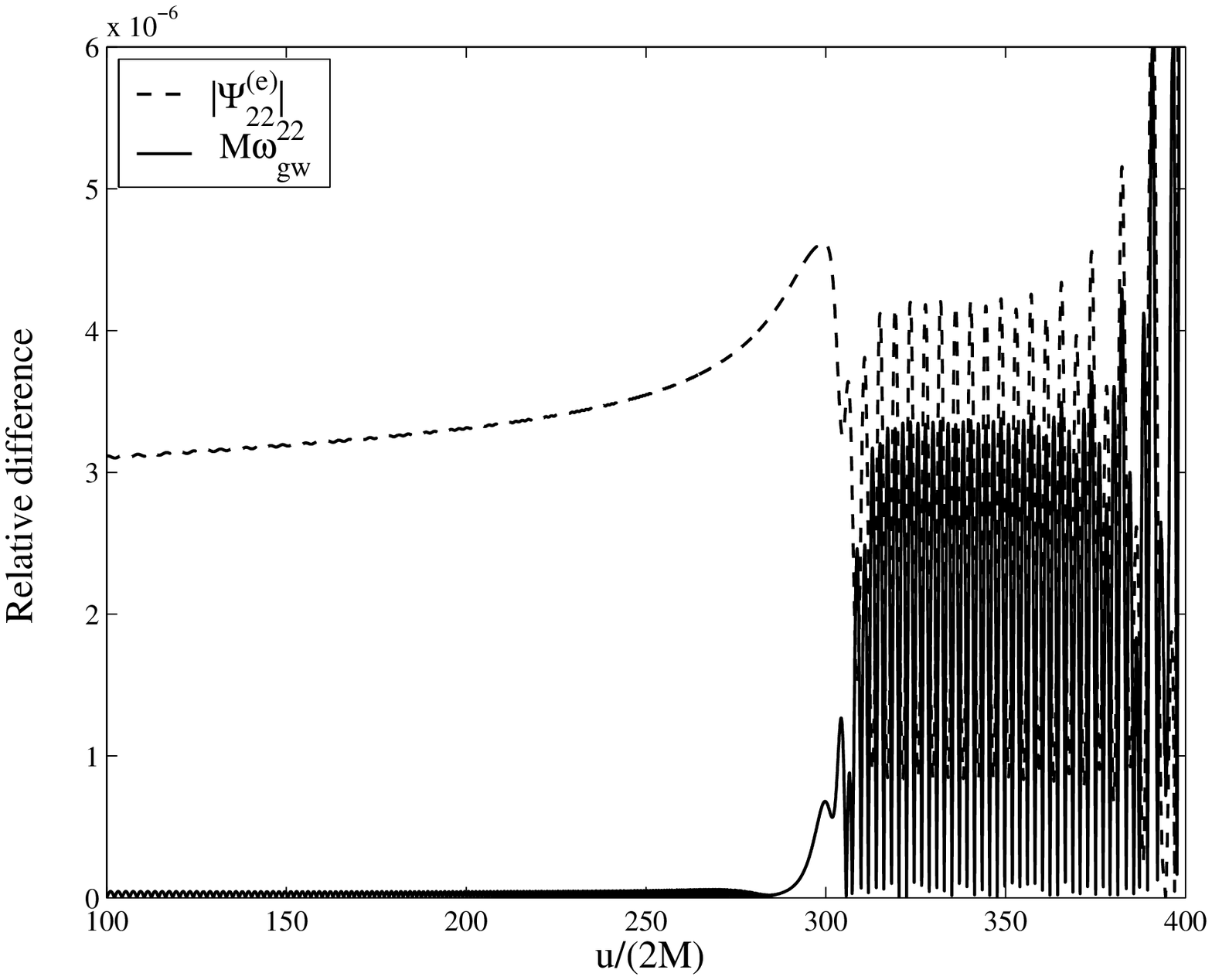}
\includegraphics[width=6.56 cm]{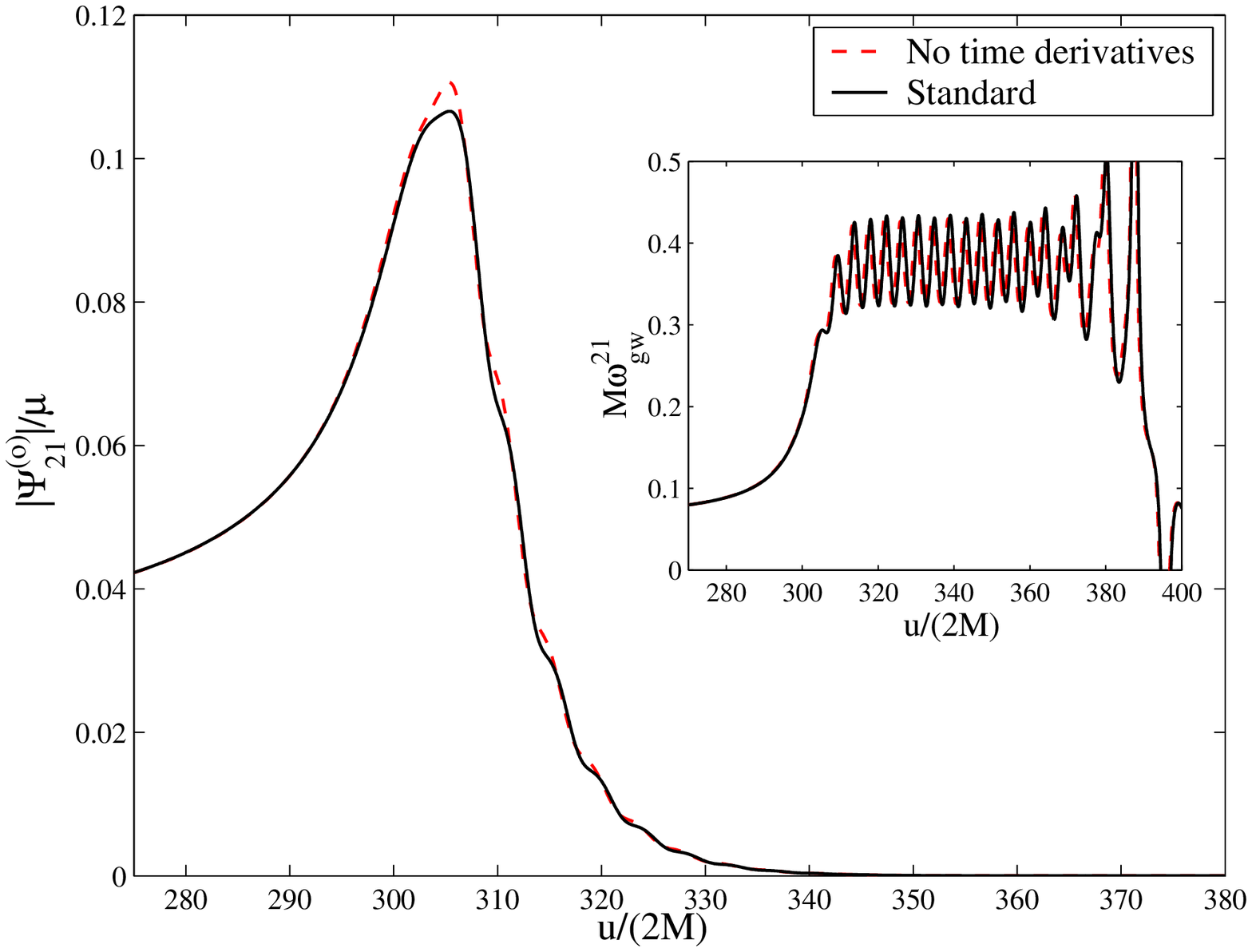}
\fl \caption{\label{fig:tests}Consistency checks of the sources. 
{\it Left panel}: relative difference between waveforms obtained with 
different expressions for $S^{(\rm e)}_{22}$.  
{\it Right panel}: check of the smallness of the effects linked to the 
terms $\propto\dot{\hat{P}}_{\varphi}$ and $\propto\dot{\hat{H}}$ in
the odd-parity source $S^{(\rm o)}_{21}$ from (\ref{src:odd}).}
\end{figure}

We conclude this section by briefly quoting two consistency checks of 
the matter source that we have performed. The first check concerns the
consistency between the two different functional forms~(\ref{source:standard}) 
or~(\ref{source:reload}) that the sources can take. As said above these
two forms would be mathematically equivalent for a real $\delta$-function,
but will differ when using the Gaussian approximation~(\ref{eq:delta}).
The left panel of figure~\ref{fig:tests} shows the relative difference 
between two gravitational wave moduli $|\Psi^{(\rm e)}_{22}|$ (dotted line): 
one of them computed using (\ref{source:standard}) and the other using 
(\ref{source:reload}) with $\Delta r_*=0.02$ and $\sigma=\Delta r_*$.
This left panel also shows the difference between the gravitational 
wave frequencies $M\omega^{22}_{\rm gw}$ (solid line) computed with the
two different sources. The relative difference~$\simeq 10^{-5}$ on 
both makes us confident that the ``smeared'' $\delta$-function is a very 
good approximation to the actual $\delta$-function. 
[Let us also comment, in passing, that we did a convergence check 
(for $\ell=m=2$) by comparing waveforms obtained by means of Gaussians 
(with $\sigma=\Delta r_*$ and $\sigma=2\Delta r_*$), finding a relative 
difference of the same order of magnitude~$\simeq 10^{-5}$].

Finally, we performed a consistency check regarding the explicit 
time derivative appearing in the coefficient of $\delta(r_*-R_*(t))$
in the odd-parity source $S^{(\rm o)}_{21}$ from (\ref{src:odd}). 
When expanded by Leibniz rule, this time derivative generates three terms. 
The term proportional to $\dot{\hat{P}}_{R_*}$ would be present along a 
geodesic motion. By contrast, the terms proportional to $\dot{\hat{P}}_{\varphi}$ 
and $\dot{\hat{H}}$ are proportional to $\nu$ and therefore vanish
in the geodesic motion limit. These last two terms generate a correction 
of order ${\cal O}(\mu){\cal O}(\nu)\propto \nu^2$ in the source.
Such terms are {\it formally} negligible in the extreme mass ratio limit 
that we consider here. In the right panel of figure~\ref{fig:tests}
we have checked to what extent these terms are {\it numerically} small.
%
%
\section{Conclusions}
\label{sec:motivs}

As we briefly mentioned in the Introduction,  beyond solving for the
first time, within a certain approximation, the problem of the plunge 
in the extreme mass ratio limit, the most important aim of this work 
was to build ``actual'' {\it numerical} waveforms to be then compared 
with {\it analytical} ones, within the EOB framework and philosophy.

\begin{figure}[t]
\begin{center}
\includegraphics[width=7.50 cm]{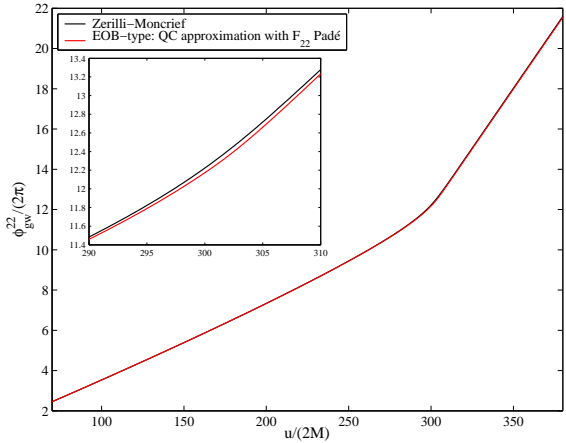}
\fl \caption{\label{fig6}Comparison between the gravitational-wave 
phase for $\ell=m=2$ obtained from the Zerilli-Moncrief equation
or from analytically matching a 3PN improved quadrupole-type formula 
(factor $F_{22}$ Pad\'e) to a superposition of quasi-normal modes.}
\end{center}
\end{figure}

We recall that the basic idea of the EOB framework is to produce
quasi-analytical waveforms by patching together a quadrupole-type
waveform during the inspiral and plunge to a QNM-type waveform 
after merger. In~\cite{DN06} we will use the numerical tools presented
here to measure to what extent these EOB-type waveforms can approximate
actual waveforms of binary black hole merger in the extreme mass ratio
limit. We give an example of this numerical-analytical comparison in 
figure~\ref{fig6}. 
This figure compares the phases of the gravitational waveforms obtained 
by two different methods, one numerical and the other semi-analytical.
In the first method, the phase has been computed by integrating in time
the instantaneous gravitational wave frequency given by (\ref{inst:wgw}),
with $\Psi^{(\rm e)}_{22}$ obtained by numerically solving the Zerilli-Moncrief
equation. The second method computes an approximate waveform in the following
way: Before crossing the light ring one uses a (Pad\'e-resummed) 3PN-improved 
quadrupole-type formula to compute the waveform from the EOB dynamics.
After crossing the light ring, the previous quadrupole-type signal, taken in
the quasi-circular (QC) approximation, is {\it matched} to a superposition of 
the first five QNMs of the black hole. Then one computes the phase of this 
matched analytical waveform by integrating the corresponding instantaneous 
gravitational wave frequency. The agreement between the ``actual'' phase and the 
``effective one body'' phase is impressively good: it turns out that the 
maximum difference between the two is less than $3\%$ of a cycle.

\ack
\label{acknw}
We are grateful to E.~Berti, A.~Buonanno, R.~De~Pietri and L.~Rezzolla for 
discussions. A.N. is supported by a post-doctoral fellowship of 
Politecnico di Torino. The computations were performed on the INFN beowulf 
cluster for Numerical Relativity \textit{Albert} at the University of Parma. 
A.N. gratefully acknowledges the support of ILIAS and IHES for hospitality 
during the completion of this work.

\section*{References}

\end{document}